# Discovery of a symmetry-driven electronic cascade in a *d*-wave altermagnet

Zhouyi Yin[1#], Zheng Shi[2#], Shuxuan Zhang[1#], Changchao Liu[3#], Xingkai Cheng[2], Fayuan Zhang[4,5], Han-Bin Deng[1], Jia-Xin Yin[1], Chaoyu Chen[4*], Guang-Han Cao[3*], Junwei Liu[2*], Yue Zhao[1,6*]

[1] State Key Laboratory of Quantum Functional Materials, Department of Physics, and Guangdong Basic Research Center of Excellence for Quantum Science, Southern University of Science and Technology, Shenzhen, China.

[2] Department of Physics, The Hong Kong University of Science and Technology, Hong Kong, China.

[3] School of Physics, Zhejiang University, Hangzhou, China.

[4] Songshan Lake Materials Laboratory, Dongguan, China.

[5] Quantum Science Center of Guangdong-Hong Kong-Macao Greater Bay Area, Shenzhen, China.

[6] Guangdong Provincial Key Laboratory of Advanced Thermoelectric Materials and Device Physics, Southern University of Science and Technology, Shenzhen, China.

[#] These authors contributed equally to this work.

[*] Correspondence should be addressed to C. Chen, G.-H. Cao, J. Liu or Y. Zhao (email: chenchaoyu@sslab.org.cn, ghcao@zju.edu.cn, liuj@ust.hk, or zhaoy@sustech.edu.cn )

## ABSTRACT

**Altermagnets host magnetic compensation together with non-relativistic spin-split bands, a coexistence enabled by crystal symmetry. Yet whether and how crystal symmetry organizes collective electronic order remains largely unexplored. Here we uncover a symmetry-driven cascade of finite-*q* charge order in a *d*-wave altermagnet $Rb_{1-\delta}V_2Te_2O$, using phase-resolved scanning tunneling microscopy. A primary density-wave instability drives an initial electronic reconstruction, followed by the emergence of a nematic component and an off-axis modulation with wave vectors geometrically tied to the preceding orders. Phase-resolved spectroscopy distinguishes these components through separate contrast-inversion energies and maps out branch-selective spectral-weight redistribution within the off-axis mode. Together with doping and temperature evolution, these observations establish a highly coordinated hierarchy of coupled density-wave instabilities, consistent with successive symmetry lowering. This multi-component hierarchy can be well described in the Landau framework through sequential softening of the charge orders, where bilinear coupling to their compatible octupolar partners at the lower-symmetry stages enables mutual stabilization within an intertwined charge-multipole state. Such a transparent realization of a charge-order cascade shows how altermagnetic symmetry can extend beyond band formation to organize collective electronic order, offering a new perspective on emergent many-body states in correlated quantum materials.**

In altermagnets, crystal symmetry enforces zero net magnetization while producing sizable spin-split band structures[1–4]. Despite intense experimental efforts to establish altermagnetism through non-relativistic single-particle band signatures[5–11], compensated magnetic configurations[12–16] and characteristic transport responses[17,18], much less is known about its collective many-body consequences. Even less clear is whether and how the symmetry principle that defines altermagnetism extends to the organization of emergent electronic order. Among various metallic altermagnetic realizations, the *d*-wave case[9,10,19] is particularly compelling because its magnetic background can be described by an ordered magnetic octupole[20]. Metallic *d*-wave altermagnets thus offer a transparent platform to explore how magnetic, orbital, and charge degrees of freedom manifest as a collective electronic landscape within the symmetry structure of altermagnetism.

Charge density waves (CDWs) provide a momentum- and phase-resolved window onto finite-*q* electronic order, resolving not only the ordering wave vectors but also the spatial contrast and energy dependence of the associated electronic reconstruction[21–24]. CDWs are frequently entangled with nematicity[25–27] and can host unconventional chiral charge textures[28], making charge order a sensitive probe of broader symmetry breaking. In most cases, their ordering wave vectors are pinned to high-symmetry directions of the parent lattice[29–33]. Spontaneous departures from these discrete orientations are rare[34] and highly informative, often pointing to nontrivial organizing constraints. Beyond their common role as primary instabilities, density waves can also emerge as intertwined or secondary responses coupled to order in other channels[35,36], leaving fingerprints in the charge sector. A pair-density wave, for example, is a finite-momentum superconducting order whose coupling to charge order induces modulations at related wave vectors[37]. These properties make CDWs a powerful diagnostic, motivating a symmetry-resolved study of finite-*q* charge order in altermagnets.

Using phase-resolved scanning tunneling microscopy (STM) on a *d*-wave altermagnet $Rb_{1-\delta}V_2Te_2O$, we uncover a hierarchical symmetry-breaking cascade of density-wave orders. The cascade begins with a primary $\boldsymbol{q}_1$ reconstruction compatible with the fourfold and mirror symmetry of the parent lattice, followed by an axis-aligned $\boldsymbol{q}_2$ component that selects a nematic direction, and culminates in an off-axis $\boldsymbol{q}_3$ modulation that departs from high-symmetry axes yet remains locked to the selected nematic domain, following the inward relation $\boldsymbol{q}_3 \approx \boldsymbol{q}_1 - \boldsymbol{q}_2$. These modulations exhibit component-specific contrast-inversion energies away from the Fermi level, while the off-axis order shows spectral inequivalence between the two parent-mirror-related branches. Together with the $\boldsymbol{q}_3$-specific amplitude and phase anomaly, this identifies $\boldsymbol{q}_3$ as a coupled but distinct off-axis channel rather than a simple geometrical consequence of the preceding orders. These observations reveal a hierarchical sequence of density-wave instabilities. Landau theory further shows that this hierarchy comes from successive softening of density-wave channels. Their bilinear coupling to compatible octupolar partners links the charge cascade to the multipolar altermagnetic background, yielding an intertwined charge-multipole state. Upon increasing electron doping or temperature, the $\boldsymbol{q}_2$ and $\boldsymbol{q}_3$ components are suppressed before the primary $\boldsymbol{q}_1$ reconstruction, defining a stability sequence aligned with the cascade.

## Multiple charge density waves on the Te-terminated surface

The unit cell of $Rb_{1-\delta}V_2Te_2O$ (Fig. 1**a**) consists of $V_2Te_2O$ layer intercalated with rubidium (Rb) atoms. Below the Néel temperature of 337 K, the neighboring vanadium moments within each layer order antiferromagnetically along $c$-axis. The resulting state is compatible with a non-relativistic spin-group description, in which spin-reversal operations combined with specific crystal symmetries ($C_{4z}/M_{1\bar{1}0}$) relate the opposite-spin sublattices[38]. This arrangement gives rise to a $d$-wave-like altermagnetic texture[9,19]. The spin-polarized bands near the Fermi energy (Fig. 1**d**) thus satisfy the symmetry constraint $\rho_{\uparrow}(\boldsymbol{k}) = \rho_{\downarrow}(C\boldsymbol{k})$ for $C \in \{C_{4z}, M_{1\bar{1}0}, M_{\bar{1}10}\}$. To access the electronic structure by STM, we prepared a clean Te-terminated surface (Extended Data Fig. 1). The surface reveals a well-ordered square lattice with in-plane lattice constant approximately 4.03 Å, in close agreement with the bulk $a$ = 4.05 Å (P4/mmm), providing a well-defined platform to investigate the electronic instabilities.

On these atomically flat terraces, the constant-current topography reveals three pronounced long-wavelength modulations superimposed on the square lattice at 4.4 K (Fig. 1**b**). First, the plaquette centers, enclosed by four surface Te atoms, alternate in height in a checkerboard arrangement, forming a commensurate $\left(\sqrt{2} \times \sqrt{2}\right)$ R45° superstructure ($\boldsymbol{q}_1$). Second, rows of Te atoms aligned with one principal crystalline axis appear periodically enhanced, showing a unidirectional, axis-aligned modulation ($\boldsymbol{q}_2$). Third, the inter-row motif displays an additional modulation, indicative of an off-axis component ($\boldsymbol{q}_3$). These textures persist across multiple fields of view and scan conditions.

To quantify the modulations, we analyze the fast Fourier transform (FFT) of the same topography (Fig. 1**c**). Using the Bragg vectors as the reciprocal-space basis, five peaks are resolved. The three dominant wave vectors are located at $\boldsymbol{q}_1 = (-0.50, 0.50)$ (commensurate $\left(\sqrt{2} \times \sqrt{2}\right)$ R45°), $\boldsymbol{q}_2 \approx (0, 0.18)$ (axis-aligned), and $\boldsymbol{q}_3 \approx (-0.50, 0.32)$(off-axis). Their electronic density wave character is supported by their non-dispersive relation over a broad bias range, distinct contrast-inversion energies, and temperature-dependent evolution discussed below. Additional satellite peaks appear near $\boldsymbol{G} - \boldsymbol{q}_2$ and $\boldsymbol{G} - 2\boldsymbol{q}_2$, consistent with the presence of the $\boldsymbol{q}_2$ modulation. The calculated constant-energy contours at the Fermi level (Fig. 1**d**) show limited overlap upon folding by $\boldsymbol{q}_{1,2,3}$, disfavoring a conventional Fermi-surface-nesting scenario.

The wave-vector star of $\boldsymbol{q}_1$ is compatible with the fourfold symmetry of the parent lattice. Within this background, the nematic axis-aligned $\boldsymbol{q}_2$ marks a rotational-symmetry breaking. In a tetragonal parent system, such axis-selection is commonly classified into two Ising-type nematic channels: the axis-aligned $B_{1g}$ and the diagonal $B_{2g}$[35,36]. The axis selection of $\boldsymbol{q}_2$ is naturally associated with the former classification. The $\boldsymbol{q}_3$ component is qualitatively distinct. It appears at an off-axis wave vector and obeys an inward geometric relation to $\boldsymbol{q}_{1,2}$ through $\boldsymbol{q}_3 \approx \boldsymbol{q}_1 - \boldsymbol{q}_2$. Such an off-axis modulation is unusual in non-chiral high-symmetry crystals, where density-wave vectors are typically locked to high-symmetry directions, except in special cases such as chiral lattices[27] or strongly correlated systems with unconventional electronic/magnetic anisotropies[34]. Its orientation, being neither axial nor diagonal, falls outside standard nematic classifications and points to a coupled finite-$q$ origin.

### Energy- and phase-resolved characterization of the off-axis $\boldsymbol{q}_3$

To further investigate the nature of the off-axis modulation $\boldsymbol{q}_3$, we perform phase- and energy-resolved Fourier analysis of differential conductance $\mathrm{d}I/\mathrm{d}V$ $(\boldsymbol{r}, E)$ maps over a dense energy grid near the Fermi level. The complex Fourier amplitudes $A(\boldsymbol{q},E) = |A(\boldsymbol{q},E)|\mathrm{e}^{\mathrm{i}\varphi(\boldsymbol{q},E)}$ provide direct access to the energy-dependent phase and intensity of individual modulations, thereby enabling different periodic components to be separated even when multiple periodicities coexist in real space[23,24].

Fig. 2**a** shows the normalized amplitude $|A(\boldsymbol{q}_3,E)|$, averaged over all $\boldsymbol{q}_3$-related peaks. Two pronounced minima occur at $E_\mathrm{H}$ ~ 14 mV and $E_\mathrm{L}$ ~ -47 mV. Across these minima, inverse-FFT-filtered maps reveal clear real-space contrast inversions of the $\boldsymbol{q}_3$-related modulation: one between 8 and 20 mV at positive bias, and the other between -58 and -35 mV at negative bias (Fig. 2**b**). Using the same Fourier-filtering convention, the line profiles show that the 20 mV and -58 mV traces are in phase, while the 8 mV and -35 mV traces are shifted by π. A weaker amplitude dip near -3 mV is not accompanied by a phase reversal and is therefore not treated as a contrast-inversion feature.

Beyond these contrast inversions, $\boldsymbol{q}_3$ exhibits an clear internal branch-inequivalence near the positive reversal. In the parent-symmetry setting, the two branches flanking the selected nematic direction, denoted $\boldsymbol{q}_3^{\pm}$, would be equivalent under the parent axial mirror reflections, yet they show a pronounced bias-dependent redistribution of spectral weight. At 20 mV, the Fourier amplitude is stronger at $\boldsymbol{q}_3^{+}$, whereas at 8 mV the amplitude reverses in favor of $\boldsymbol{q}_3^{-}$ with a diminishing $\boldsymbol{q}_3^{+}$ (Fig. 2**c** and **d**). Importantly, the diffuse FFT background shows no corresponding directional imbalance across these energies, inconsistent with tip-induced anisotropy or trivial instrumental asymmetry (Extended Data Fig. 2). The $\boldsymbol{q}_3$ peaks are selectively resolved on the inward side, consistent with $\boldsymbol{q}_3 \approx \boldsymbol{q}_1 - \boldsymbol{q}_2$, without a comparable outward $\boldsymbol{q}_1 + \boldsymbol{q}_2$ counterpart. The branch-dependent redistribution between the mirror-related $\boldsymbol{q}_3^{\pm}$, together with this inward selectivity, provides a symmetry-sensitive signature of a lower-symmetry off-axis response, with $\boldsymbol{q}_3$ emerging as a spectroscopically resolved channel coupled to the coexisting $\boldsymbol{q}_1$ and $\boldsymbol{q}_2$ modulations.

## Phase-resolved hierarchy of the CDW orders

Motivated by the branch-resolved off-axis response of $\boldsymbol{q}_3$, we next examine the energy-dependent evolution of all three modulations to clarify how they are related within the multi-component density-wave state. Fig. 3**a** plots the complex Fourier phase $\varphi(\boldsymbol{q},E)$ for $\boldsymbol{q}_{1,2,3}$ extracted from energy-resolved d*I*/d*V* maps.

The $\boldsymbol{q}_1$ component exhibits a single, sharp π phase jump around -56 mV (Fig. 3**a**), consistent with the contrast inversion observed in the inverse-FFT-filtered real-space maps (Extended Data Fig. 3). No additional full phase reversal is detected over the measured energy range, indicating that $\boldsymbol{q}_1$ is governed by a single dominant electronic reconstruction. To assess its band-structure origin, we perform an energy-resolved spin-conserving band-overlap analysis for the three CDW wave vectors (Fig. 3**c**; Methods). The overlap response shows a pronounced finite-binding-energy enhancement near -46 mV for $\boldsymbol{q}_1$, close to the experimentally observed $\boldsymbol{q}_1$ contrast-inversion energy near -56 mV. This enhancement arises from substantial *β*-*γ* overlap between folded X- and Y-pocket contours (Extended Data Fig. 4**e**, inset). By contrast, no comparable peak is found for $\boldsymbol{q}_2$ or $\boldsymbol{q}_3$.

The subsidiary components exhibit a richer phase structure. The $\boldsymbol{q}_2$ component displays two near-$\pi$ phase reversals, a relative sharp one near -56 mV and a broad cumulative phase rotation around 0 mV, each accompanied by a real-space contrast inversion of the unidirectional stripe modulation (Extended Data Fig. 3). The -56 mV reversal coincides with the $\boldsymbol{q}_1$ reconstruction scale and may reflect coupling to the primary order, whereas the near-$E_F$ reversal has no corresponding full contrast inversion in either $\boldsymbol{q}_1$ or $\boldsymbol{q}_3$, supporting its identification as a spectroscopically distinct nematic channel.

The $\boldsymbol{q}_3$ component undergoes two phase reversals near -47 mV and 14 mV, close to but not coincident with the $\boldsymbol{q}_1$ and $\boldsymbol{q}_2$ contrast-inversion windows. The positive-bias anomaly is particularly informative: the branch-resolved $\boldsymbol{q}_3^{\pm}$ phases reverse within an energy range where both $\boldsymbol{q}_1$ and $\boldsymbol{q}_2$ phases remain relatively stable. Over this range, the $\boldsymbol{q}_3^{\pm}$ Fourier amplitudes are strongly suppressed while the $\boldsymbol{q}_1^{\pm}$ and $\boldsymbol{q}_2$ amplitudes stay finite or continue to increase (Extended Data Fig. 5). A purely passive $\boldsymbol{q}_1 - \boldsymbol{q}_2$ mixing picture would instead be expected to scale with the parent amplitudes and inherit the phase relation $\varphi_1 - \varphi_2$. The simultaneous $\boldsymbol{q}_3$-specific amplitude suppression and phase reversal therefore cannot be reduced to simple passive mixing. Combined with the dominant branch switching shown in Fig. 2, this supports the assignment of $\boldsymbol{q}_3$ as a coupled but distinct lower-symmetry off-axis channel with branch-resolved inequivalence. Moreover, in domains where the $\boldsymbol{q}_2$ nematic axis rotates, the split $\boldsymbol{q}_3^{\pm}$ branches reorient in tandem (Extended Data Fig. 6), suggesting that the off-axis pair is tied to the local nematic domain.

Together, the phase-resolved spectra, domain-locking and band-overlap analysis identify a hierarchy in which $\boldsymbol{q}_1$ forms the primary density-wave reconstruction, $\boldsymbol{q}_2$ defines the subsidiary axis-aligned nematic channel, and $\boldsymbol{q}_3$ emerges as a branch-inequivalent off-axis component locked to the nematic domain. This experimentally established hierarchy provides the basis for the symmetry-organized cascade discussed below.

**Landau description of the CDW cascade**

To formulate this hierarchy in symmetry terms, we introduce three coupled order-parameter channels corresponding to each CDW component. In this Landau description, the symmetry labels refer to the parent-symmetry classification of the coupled order-parameter channels. The primary order parameter $Q_1$ at $\boldsymbol{q}_1$ is compatible with the parent $D_{4\mathrm{h}}$ structure and initiates an electronic reconstruction. Within this reconstructed background, the subsidiary $\boldsymbol{q}_2$ order is described by an effective nematic variable $N_2$, which selects an axis-aligned stripe orientation and reduces the point group to $D_{2\mathrm{h}}$. The subsequent off-axis component is described by $N_3$, which brings the inequivalence between the parent-mirror-related $\boldsymbol{q}_3^{\pm}$ branches and further drives the system into a $C_{2\mathrm{h}}$ state. The resulting subgroup chain, $D_{4\mathrm{h}} \supset D_{2\mathrm{h}} \supset C_{2\mathrm{h}}$, captures the progressive symmetry descent and provides a phenomenological description of the CDW cascade.

We first consider the charge sector using a minimal phenomenological free energy:

$$F_{1,\mathrm{CDW}} \sim \alpha_1 (T - T_1) {Q_1}^2 + \beta_1 {Q_1}^4$$
$$F_{2,\mathrm{CDW}} \sim (\alpha_2 - \gamma_2 Q_1^2) N_2^2 + \beta_2 N_2^4$$
$$F_{3,\mathrm{CDW}} \sim (\alpha_3 - \gamma_3 N_2^2) N_3^2 + \beta_3 N_3^4$$

Here $N_2 = |Q_2^x| - |Q_2^y|$ measures the imbalance between the two possible stripe orientations, while $N_3 = |Q_3^+| - |Q_3^-|$ evaluates the branch inequivalence of the off-axis order. This form encodes a sequential softening cascade. Supported by the finite-binding-energy band-overlap channel identified above, the primary order parameter $Q_1$ condenses below $T_1$ and initiates the leading electronic reconstruction. As $Q_1^2$ grows, it reduces the effective quadratic coefficient of $N_2$. Once it goes beyond the critical threshold ($Q_1^2 = \alpha_2/\gamma_2$), the quadratic coefficient of the nematic channel changes sign, allowing a finite $N_2$ to develop (Extended Data Fig. 7). Following the same mechanism, the growth of $N_2$ subsequently lowers the effective free-energy cost of $N_3$, promoting the off-axis branch-splitting order.

In a *d*-wave altermagnet, the electronic reconstruction is naturally coupled with the underlying magnetic multipolar background. Magnetic octupolar order has been identified as an order-parameter description of *d*-wave altermagnetism[20]. The symmetry descent generated by the CDW cascade therefore allows matching octupolar channels of the altermagnetic background to couple bilinearly to the charge hierarchy. Denoting the octupolar channels compatible with $N_\mathrm{i}$ as $M_\mathrm{i}$ ($\mathrm{i} = 2,3$), the leading symmetry-allowed free-energy terms can be written as

$$F_{2,\mathrm{OCT}} \sim F_{2,\mathrm{CDW}} + r_2 M_2^2 + u_2 M_2^4 + \lambda_2 M_2 N_2$$
$$F_{3,\mathrm{OCT}} \sim F_{3,\mathrm{CDW}} + r_3 M_3^2 + u_3 M_3^4 + \lambda_3 M_3 N_3.$$

A finite $N_\mathrm{i}$ can therefore induce $M_\mathrm{i}$. Relaxing over this induced response yields an effective negative contribution, $\Delta F_\mathrm{i} \sim -\lambda_\mathrm{i}^2 N_\mathrm{i}^2/4r_\mathrm{i}$, which reduces the effective free-energy cost of the $N_\mathrm{i}$ channel (Supplementary Information). This coupling provides an energetic route that favors the downstream emergence of $\boldsymbol{q}_2$ and $\boldsymbol{q}_3$, despite the absence of comparable bare-band overlap enhancements at their wave vectors.

The cascade therefore reshapes the hidden octupolar background. In the parent $D_{4\mathrm{h}}$ altermagnetic state, crystal symmetries enforce the compensation between spin-opposite sublattice, giving rise to the octupolar configuration schematically shown in Fig. 3**d**. As the charge hierarchy develops, additional octupolar components become allowed and are induced by the corresponding charge variables (Fig. 3**b**). Their superposition redistributes the octupolar weight, producing a distorted lower-symmetry texture, in which a secondary net-moment-like imbalance can also emerge (Fig. 3**e**). This evolution illustrates how the finite-$q$ charge hierarchy and the hidden multipolar background become intertwined through symmetry-allowed coupling. The $N_3$ channel, manifested experimentally as the off-axis modulation, provides a particularly sensitive real-space signature of the resulting multipolar imprint, as it emerges only at the final, mirror-breaking stage of the cascade.

**Doping and temperature dependence**

We next examine the stability of this cascade by tracking the doping and temperature evolution of the three modulations. In such a cascade, weakening the primary reconstruction is expected to sequentially suppress the lower-symmetry components. To access doping dependence, we compare spatially distinct regions with different local electron doping. These regions are identified by quasiparticle-interference (QPI) measurements, using the energy dispersion of a characteristic scattering wave vector as a local Fermi-level marker (Extended Data Fig. 4). The three-$q$ coexisting area discussed above is used as the reference, denoted as region I. Relative to region I, region I' exhibits a decreased electron doping with a downward Fermi-level shift of ~ 15 meV, whereas regions II and III exhibit upward Fermi-level shifts of ~ 25 and ~ 34 meV, respectively, corresponding to progressively stronger electron doping.

At 4.4 K, the normalized Fourier spectra reveal a sequential suppression of the CDW components with increasing electron doping (Fig. 4**b**, **c**). From region I to region II, the $\boldsymbol{q}_2$ and $\boldsymbol{q}_3$ peaks are selectively suppressed, while the $\boldsymbol{q}_1$ checkerboard remains visible but weakened. In the most electron-doped region III, all superlattice peaks disappear. The corresponding real-space images show the same progression (Fig. 4**e-g**). This doping-dependent suppression sequence is consistent with the hierarchy established above. It is also consistent with the finite-binding-energy band-overlap picture, in which increasing electron doping shifts the relevant $\boldsymbol{q}_1$ overlap channel farther from the Fermi level and weakens the primary reconstruction.

Temperature-dependent measurements further support this hierarchy. Representative $d^2I/dV^2$ spectra measured in region I shows that a weak should-like feature near -46 mV, close to the $\boldsymbol{q}_1$ contrast-inversion energy, despite substantial smearing of the positive-bias features at 77 K (Fig. 4**a** inset). In the relatively hole-shifted region I', where the Fermi level lies closer to the finite-binding-energy overlap condition, the $\boldsymbol{q}_1$ peak remains pronounced at 77 K (Fig. 4**b**, **c**). Together with the doping-dependent suppression sequence, these observations indicate that the primary checkerboard reconstruction is the most robust component of the cascade, with the nematic and off-axis channels confined to more restricted stability windows.

In summary, our real-space and spectroscopic measurements establish that the observed density-wave orders form a coupled charge-order cascade rather than a collection of unrelated Fourier components. The component-specific contrast-inversion energies, branch-dependent off-axis response, domain locking between the nematic and off-axis components, and systematic doping and temperature evolution show that the subsidiary orders are spectroscopically resolved lower-symmetry channels within the primary reconstructed background. This organization is captured by a Landau framework in which successive charge channels soften and couple to compatible octupolar components, linking the hierarchy to an intertwined charge-multipole state.

Electronic cascades have been observed in moiré flat bands[39] and kagome metals[33], where scanning tunneling measurements revealed a series of reconstructed electronic states. The cascade resolved here has a distinct implication: it is a hierarchy of finite-$q$ charge order parameters constrained by the multipolar structure of the parent $d$-wave altermagnet. Through symmetry-allowed charge-multipole coupling, the otherwise hidden octupolar constraints become experimentally legible in real space. The charge cascade thus carries an imprint of the altermagnetic multipolar texture, motivating future spin-resolved probes to test whether the density waves are accompanied by corresponding spin-sensitive or multipolar signatures. More broadly, our results provide a transparent real-space realization of altermagnetic symmetry as an organizing principle for emergent collective states, offering a perspective for disentangling complex symmetry-organized phases in correlated quantum materials.

## Figures

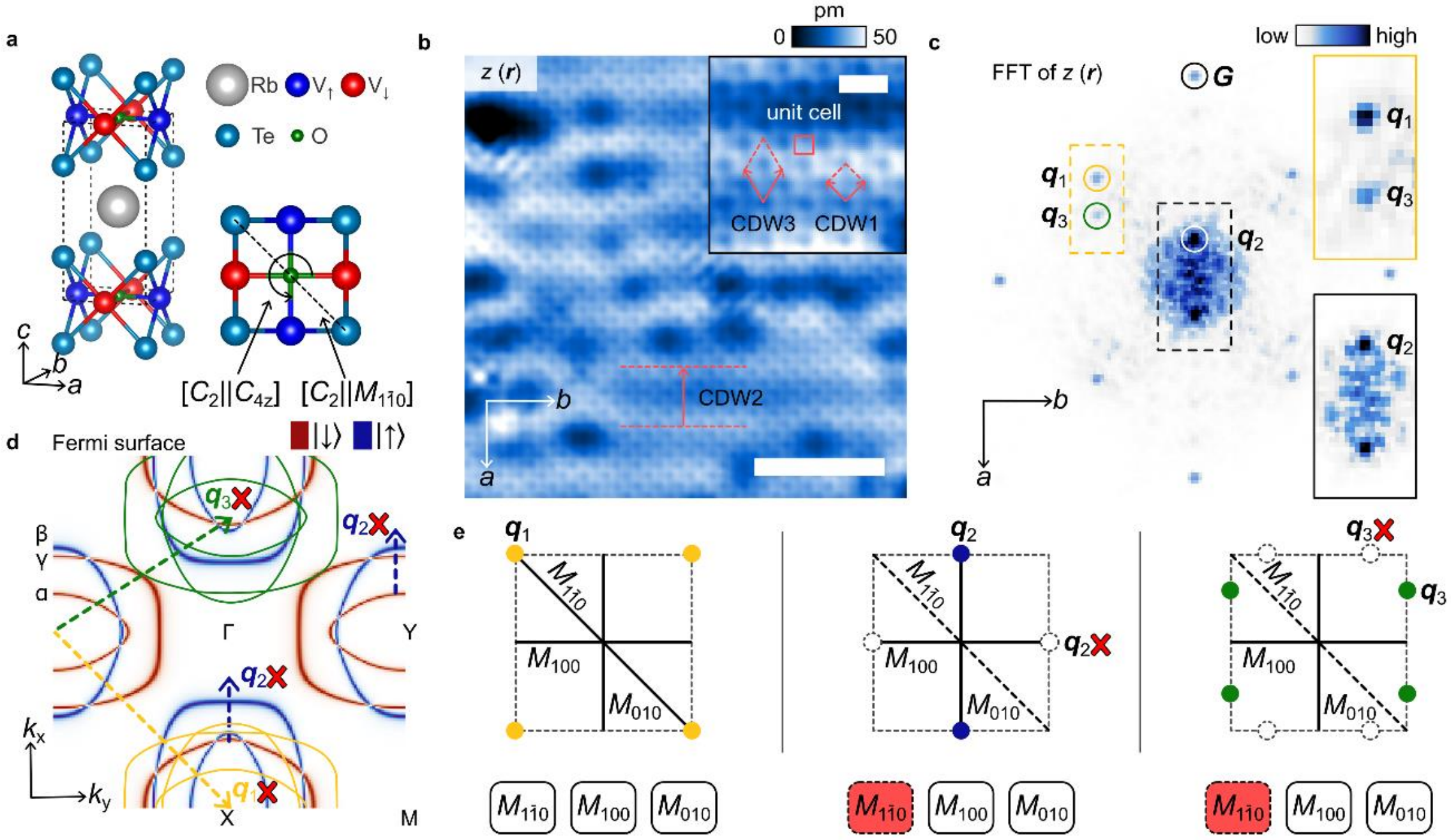

**Figure 1. Three charge-density-wave components and their symmetry characteristics on the Te-terminated surface.**

**a,** Crystal structure of $Rb_{1-\delta}V_2Te_2O$. In the top view of the $V_2Te_2O$ layer, vanadium atoms satisfy composite symmetries $[C_2\|C_{4z}]$, $[C_2\|M_{1\bar{1}0}]$ ($[C_2\|M_{\bar{1}10}]$). **b,** Topographic image of the Te-terminated surface, showing three charge-density-wave modulations, denoted as CDW1, CDW2, and CDW3. The red box in the inset outlines the Te unit cell. Setpoint: $V_{\text{bias}} = 60$ mV, $I_t = 50$ pA. Scale bars: 5 nm (main panel), 1 nm (inset). **c,** FFT of **b**, showing Bragg peaks $\boldsymbol{G}$ and CDW wavevectors $\boldsymbol{q}_1$, $\boldsymbol{q}_2$ and $\boldsymbol{q}_3$, corresponding to CDW1, CDW2, and CDW3, respectively. The enlarged insets use independent color scales. **d,** Spin-resolved Fermi-surface contours and schematic folding check. Translating the contours by $\boldsymbol{q}_1$ (yellow) and $\boldsymbol{q}_3$ (green) produces negligible overlap between the folded and original bands. The shorter vector $\boldsymbol{q}_2$ (blue dashed arrow) does not connect obvious parallel Fermi-surface segments, arguing against a conventional Fermi-surface-nesting origin. **e,** Symmetry properties of the three CDW orders. The $\boldsymbol{q}_1$ wave vectors are related by the mirror operations of the parent lattice. The stripe-like $\boldsymbol{q}_2$ selects one crystallographic axis, defining an axis-aligned nematic state. The off-axis $\boldsymbol{q}_3$ components appear in the inward $\boldsymbol{q}_1 - \boldsymbol{q}_2$ sector selected by the nematic domain.

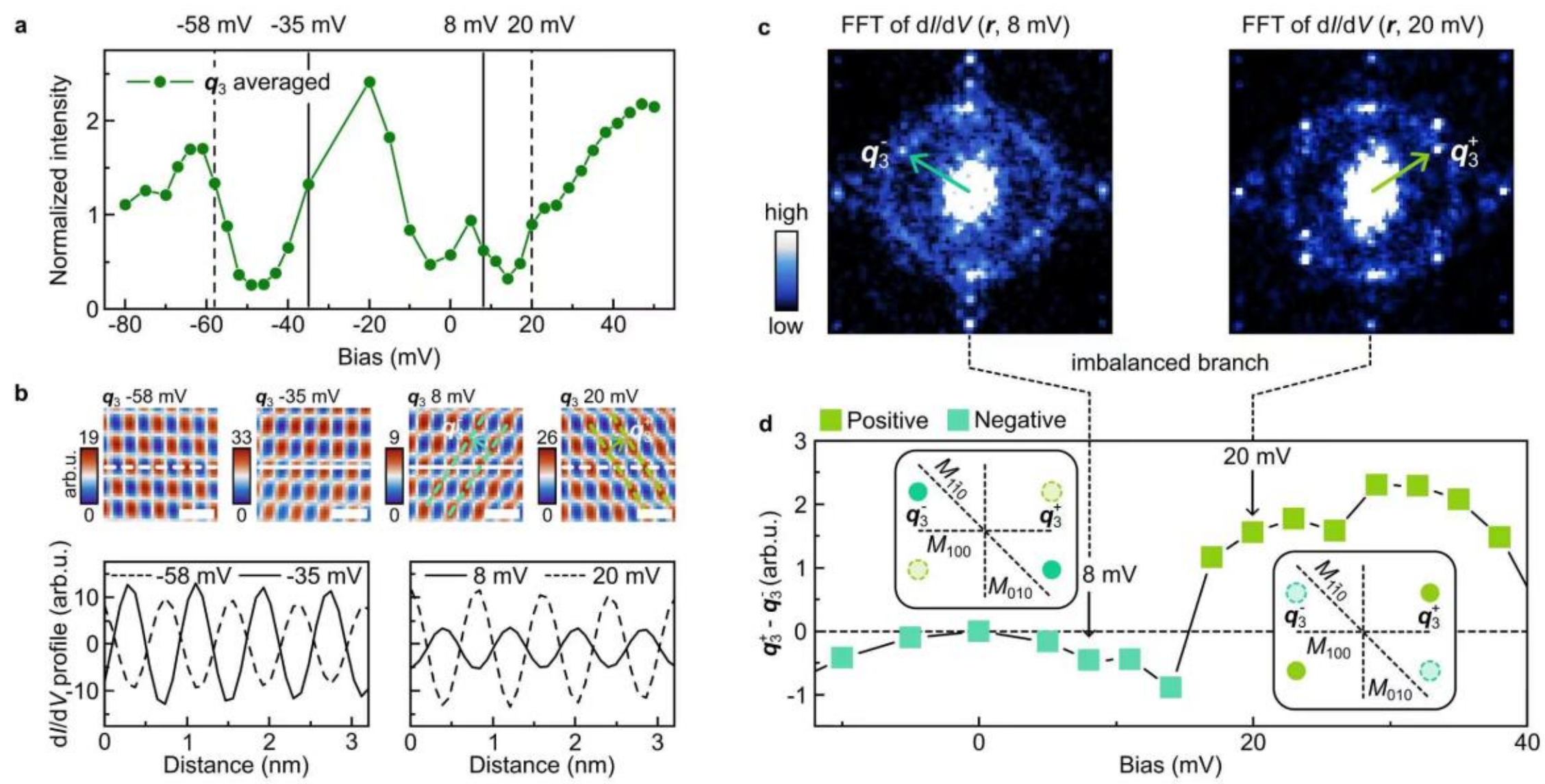

**Figure 2. Energy-dependent contrast inversion and branch inequivalence of the off-axis $\boldsymbol{q}_3$. a,** Bias-dependent normalized amplitude of the $\boldsymbol{q}_3$ component, obtained by summing over all $\boldsymbol{q}_3$-related Fourier peaks and normalizing by the total Bragg-peak amplitude. **b,** Inverse-FFT-filtered real-space images isolating the $\boldsymbol{q}_3$ modulation (top), with corresponding line profiles at -58 mV, -35 mV, 8 mV and 20 mV (bottom). The $\boldsymbol{q}_3$ modulation undergoes contrast inversion in two distinct energy windows, from -58 to -35 mV and from 8 to 20 mV, as evidenced by the corresponding $\pi$ phase shifts. These inversions identify spectral reconstruction energies near -47 mV and 14 mV. At 8 mV and 20 mV, the $\boldsymbol{q}_3$ modulation is dominated by the $\boldsymbol{q}_3^-$ branch (blue-green arrows) and $\boldsymbol{q}_3^+$ branch(yellow-green arrows), respectively. Scale bars: 1 nm. Setpoints: $V_{\text{bias}}$ = -80 mV, $I_{\text{t}}$ = 1 nA (-58 mV and -35 mV); $V_{\text{bias}}$ = -50 mV, $I_{\text{t}}$ = 500 pA (8 mV and 20 mV). **c,** FFT of d$I$/d$V$ maps at 8 mV and 20 mV. The dominant spectral weight switches between the two mirror-related off-axis branches $\boldsymbol{q}_3^-$ and $\boldsymbol{q}_3^+$, revealing an energy-dependent branch imbalance. Setpoint: $V_{\text{bias}}$ = -50 mV, $I_{\text{t}}$ = 500 pA. **d,** Difference in Fourier weight between $\boldsymbol{q}_3^+$ and its mirror-related counterpart $\boldsymbol{q}_3^-$, quantifying the branch-resolved inequivalence near the positive-bias $\boldsymbol{q}_3$ minimum.

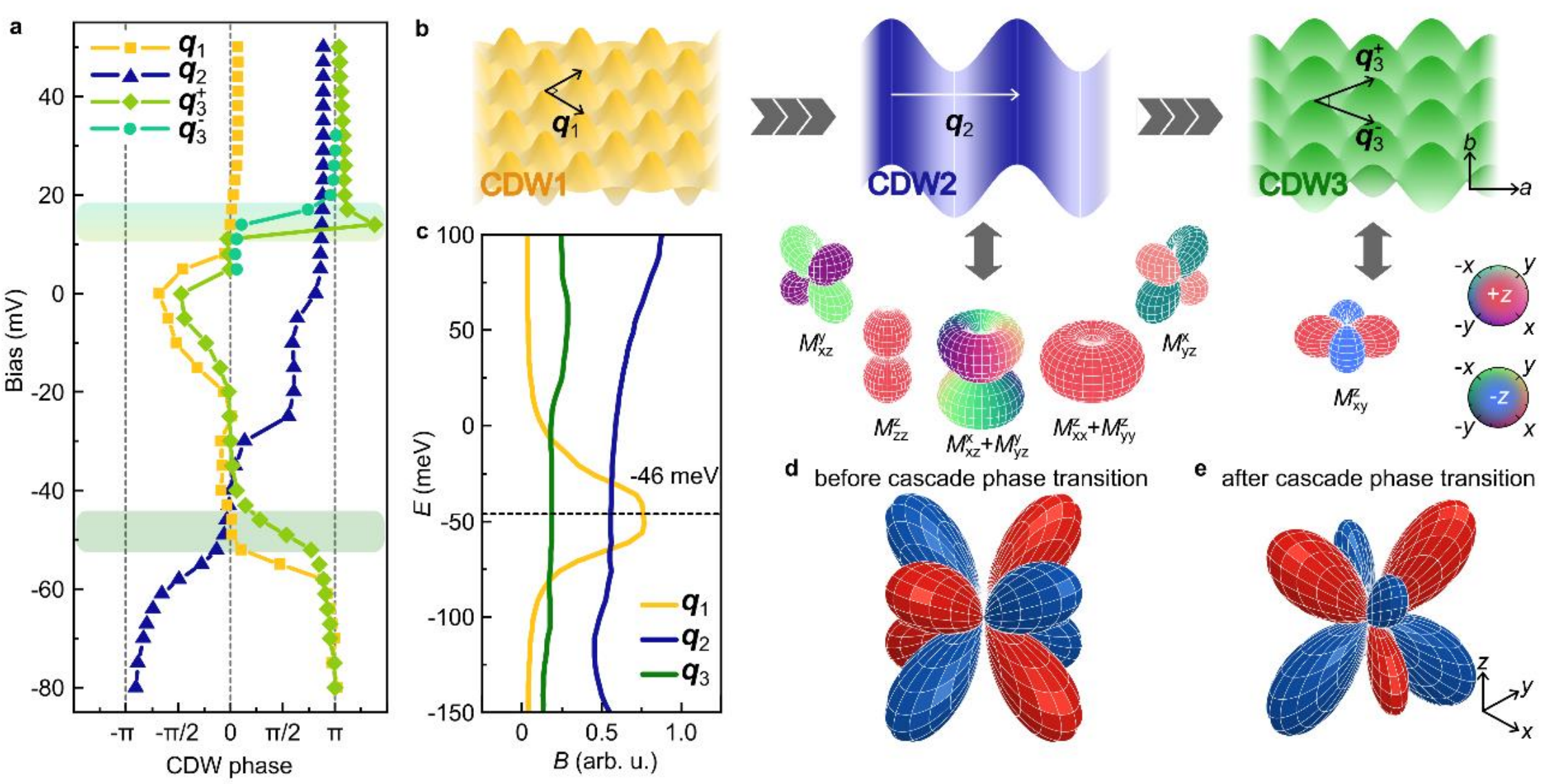

**Figure 3. Phase-resolved hierarchy and intertwined electronic-multipolar reconstruction.**
**a,** Energy-dependent phase of $\boldsymbol{q}_1$, $\boldsymbol{q}_2$, $\boldsymbol{q}_3^+$ and $\boldsymbol{q}_3^-$. Shaded regions mark π-phase shifts associated with real-space contrast inversions and characteristic reconstruction energies. Phase traces are shifted by constant offsets for visual comparison. The $\boldsymbol{q}_3^-$ trace is shown only in the positive-bias phase-reversal window. **b,** Schematic illustration of the cascade and the coupling between the lower-symmetry CDW channels and compatible octupolar components. The three-dimensional surfaces represent octupolar components allowed by the symmetry of the reconstructed CDW states, with the lobe amplitude denoting the magnitude of the magnetization-density component and the color indicating the local spin orientation. **c,** Energy-resolved spin-conserving band-overlap $B(\boldsymbol{q}, E)$, calculated from the density functional theory (DFT) band structure in the absence of CDW reconstruction for $\boldsymbol{q}_1$, $\boldsymbol{q}_2$ and $\boldsymbol{q}_3$. A pronounced enhancement near -46 meV coincides with the contrast-inversion energy of $\boldsymbol{q}_1$, whereas no comparable enhancement is observed for $\boldsymbol{q}_2$ or $\boldsymbol{q}_3$. **d**,**e,** Schematic magnetic multipolar configurations before (**d**) and after (**e**) the hierarchical CDW reconstruction, illustrating the Landau-inferred symmetry lowering and redistribution of multipolar components.

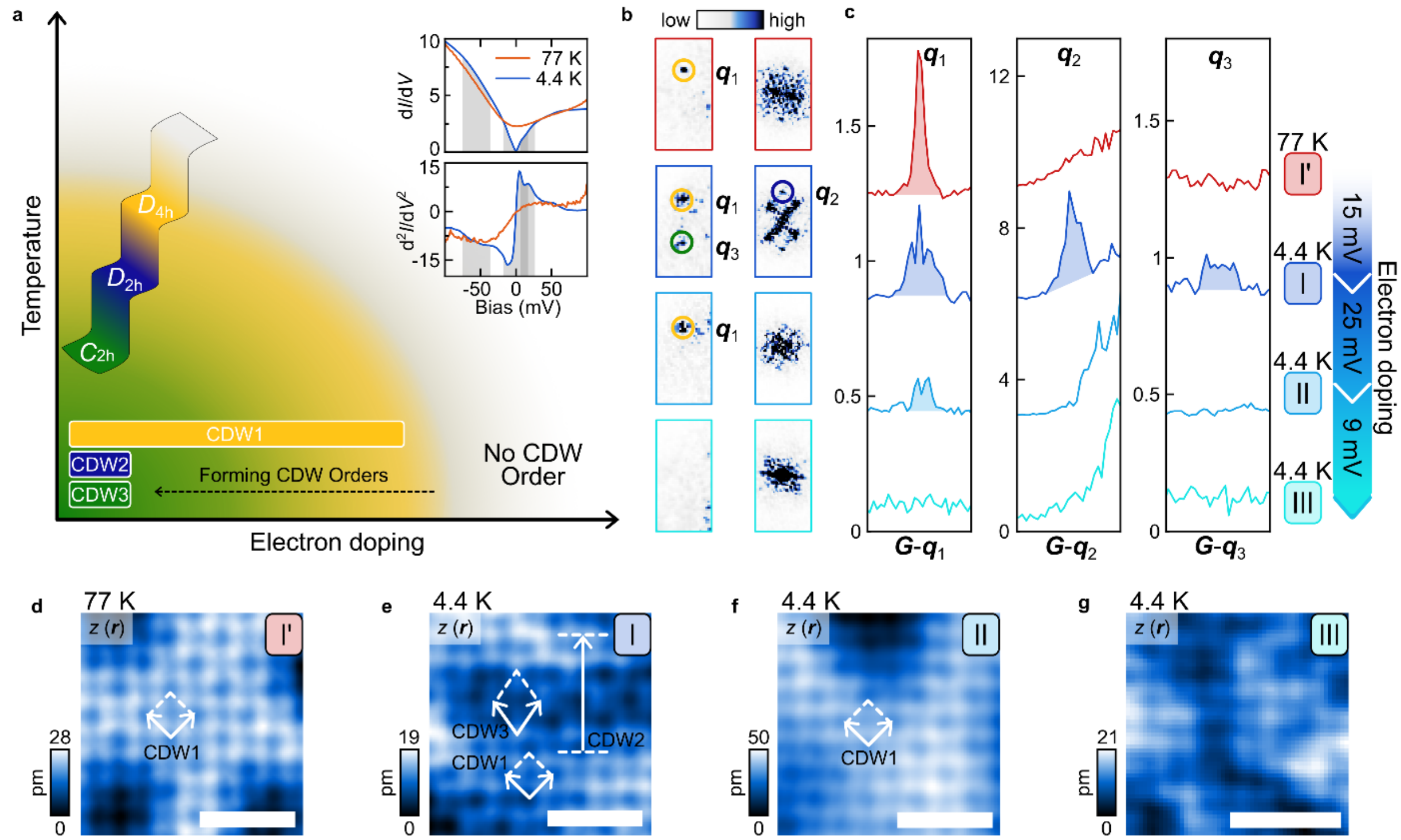

**Figure 4. Doping- and temperature-dependent evolution of the hierarchical CDW orders.** **a,** Schematic phase diagram summarizing the evolution of CDW orders with electron doping and temperature. Inset: tunneling spectra, d$I$/d$V$, and their derivative, d$^2I$/d$V^2$, measured at 4.4 K and 77 K. Grey shaded regions mark the contrast-inversion energies of the CDW modulations. Setpoint: $V_{\mathrm{bias}}$ = -150 mV, $I_{\mathrm{t}}$ = 1 nA. **b,** Cropped FFTs of topographic images acquired, from top to bottom, in region I' at 77 K, and in region I, II and III at 4.4 K. Corresponding setpoints are $V_{\mathrm{bias}}$ = 100, 25, 20, and 40 mV, and $I_{\mathrm{t}}$ = 50, 100, 50, and 50 pA. **c,** Normalized amplitude profiles of the $\boldsymbol{q}_1$, $\boldsymbol{q}_2$ and $\boldsymbol{q}_3$ components, extracted along the corresponding $\boldsymbol{G}-\boldsymbol{q}_i$ directions from the FFTs in **b**. Intensities are normalized to their corresponding Bragg-peak intensities. **d-g,** Topographic images corresponding to the FFTs in **b**. The $\boldsymbol{q}_1$-related $(\sqrt{2}\times\sqrt{2})R45°$ superstructure remains visible at 77 K in region I' (**d**). At 4.4 K, all three CDW components coexist in region I (**e**), evolve into a predominantly $\boldsymbol{q}_1$-related superstructure in region II (**f**), and are strongly suppressed in region III (**g**). Scale bars: 2 nm.

## STM measurements and surface preparation

STM measurements were performed using a Unisoku USM-1500 scanning probe microscope system. Differential conductance d*I*/d*V* spectra and maps were acquired with a Nanonis controller using a built-in lock-in amplifier at a modulation frequency of 971.9 Hz. Single crystals of $Rb_{1-\delta}V_2Te_2O$ were grown and characterized following the procedures reported in our previous work[9].

$Rb_{1-\delta}V_2Te_2O$ samples were cleaved at 77 K under ultrahigh-vacuum. The as-cleaved surface is initially covered by reconstructed Rb atoms and clustered debris. These Rb atoms are weakly bound to the underlying $V_2Te_2O$ layer and can be displaced by the STM tip. Under tunneling conditions of $V_{bias}$ = 10 mV and $I_t$ = 2 nA, repeated scanning removes both the reconstructed Rb atoms and clustered debris, pushing them toward the lower edge of the scan region. This procedure exposed a Te-terminated surface suitable for high-resolution STM measurements, as illustrated in Extended Data Fig. 1.

## Origin of passive wave-vector mixing and validation of intrinsic CDW components

Wave vector mixing can arise when multiple CDW components coexist and contribute nonlinearly to the measured electronic contrast[40,41]. We describe each CDW component by a complex Fourier coefficient $\tilde{\rho}_i(E) = A_i(E)\mathrm{e}^{\mathrm{i}\varphi_i(E)}$, so that the real-space modulation contains $\tilde{\rho}_i \mathrm{e}^{\mathrm{i}\,\boldsymbol{q}_i\cdot\boldsymbol{r}} + \tilde{\rho}_i{}^* \mathrm{e}^{-\mathrm{i}\,\boldsymbol{q}_i\cdot\boldsymbol{r}}$. For two CDW orders $\tilde{\rho}_i$ and $\tilde{\rho}_j$, the bilinear products $\tilde{\rho}_i\tilde{\rho}_j{}^*$ and $\tilde{\rho}_i\tilde{\rho}_j$ generate secondary Fourier weight at $\boldsymbol{q}_i - \boldsymbol{q}_j$ and $\boldsymbol{q}_i + \boldsymbol{q}_j$, respectively. Both terms have amplitudes proportional to $A_iA_j$, with their phases following $\varphi_i - \varphi_j$ and $\varphi_i + \varphi_j$, apart from constant phase offsets absorbed into the complex mixing coefficients.

For the experimentally relevant relation $\boldsymbol{q}_3 \sim \boldsymbol{q}_1 - \boldsymbol{q}_2$, a passive mixing contribution to the $\boldsymbol{q}_3$ Fourier coefficient can be written as $\tilde{\rho}_3{}^{mix}(E) \sim \lambda\tilde{\rho}_1(E)\tilde{\rho}_2{}^*(E)$, where $\lambda$ is a generally complex coefficient accounting for nonlinear contrast generation, tunneling form factors and Fourier-filtering conventions. In a purely passive mixing picture, $\lambda$ is expected to vary smoothly with energy over the local energy window of interest. The corresponding passive mixing has its amplitude and phase: $A_3{}^{mix}(E) \sim |\lambda| A_1(E)A_2(E)$, $\varphi_3{}^{mix}(E) \sim \varphi_1(E) - \varphi_2(E) + arg\lambda$. A purely passive $\boldsymbol{q}_1 - \boldsymbol{q}_2$ mixing signal should follow the parent amplitude product and inherit the parent phase difference, up to a smooth complex conversion factor.

We therefore tested the passive-mixing scenario by comparing the measured $\boldsymbol{q}_3$ amplitude and phase with the parent-channel product $A_1A_2$, and phase differences of $\varphi_1 - \varphi_2$. The $q_3$-specific amplitude suppression and phase reversal at positive bias, occurring while the parent $q_1$ and $q_2$ channels remain finite and phase-stable (Extended Data Fig. 5), cannot be accounted for by passive mixing with a smooth $\lambda$.

## Energy-resolved spin-conserving band-overlap analysis

Based on the DFT-derived band structure, we evaluated an energy-resolved spin-conserving band-overlap function,

$$B(\boldsymbol{q}, E) = \sum_{s=\uparrow,\downarrow} \int_{BZ} \rho_s(\boldsymbol{k} + \boldsymbol{q}, E)\rho_s(\boldsymbol{k}, E) d^2k.$$

$\rho_s(\boldsymbol{k}, E)$ is the spin-resolved spectral weight. Given the spin-polarized nature of the altermagnetic bands, the summation is restricted to spin-conserving channels with the same spin projection. This quantity measures the constant-energy phase space for connecting spin-compatible states separated by $\boldsymbol{q}$. It is therefore used as a geometric band-overlap diagnostic, rather than as a full susceptibility calculation. A peak in $B(\boldsymbol{q}, E)$ identifies a possible finite-binding-energy band-overlap channel that can contribute to the primary $\boldsymbol{q}_1$ reconstruction. Conversely, the absence of comparable peaks at $\boldsymbol{q}_2$ and $\boldsymbol{q}_3$ shown in Fig. 3**c** does not exclude phonon-assisted[21] or interaction-driven[35,42] effects that may help stabilize the subsidiary components. The involvement of such microscopic mechanisms would remain compatible with their downstream role in the symmetry hierarchy.

**Fermi level determination and local doping calibration**

The local Fermi level alignment was determined by comparing QPI patterns with the density-functional-theory (DFT) band structure, following the procedure described in our previous work[9]. In region III, the QPI image displays a clear arc-like feature, $\boldsymbol{q}_{arc}$, which matches the calculated spin-selective scattering probability of the γ band (Extended Data Fig. 4**a**). The fitted dispersion of $\boldsymbol{q}_{arc}$ along the Γ-Y direction agrees with the shifted DFT γ-band dispersion, indicating that the local Fermi level in region III corresponds to approximately -205 meV in the DFT energy reference[9] (Extended Data Fig. 4**b**). This shift is likely associated with the removal of surface Rb atoms during sample preparation, while additional local doping variations can arise from defects and inhomogeneous Rb intercalation.

Relative doping between different regions was calibrated using a characteristic scattering wave vector $\boldsymbol{q}_\delta$. In region I, the QPI pattern at -140 mV exhibits four prominent $\boldsymbol{q}_\delta$ features, consistent with intra-δ-band scattering from intercalated-Rb-derived electronic states (Extended Data Fig. 4**c)**. Because $\boldsymbol{q}_\delta$ is resolved across all characterized regions over a broad energy range, it provides an internal marker of relative local doping. The $2\boldsymbol{q}_\delta$ dispersion along the Γ-M direction were fitted using a common dispersion slope referenced to the region-I data (Extended Data Fig. 4**e**). The extracted energy offsets define the relative doping values used in Fig. 4.

**Methods references**

**Acknowledgements**

This work is supported by the National Key R&D Program of China (grant nos. 2022YFA1403700 to Y.Z. and C.C., 2023YFA1406101 to G.-H.C., 2025YFA1411200 to C.C. and 2025YFA1411500 to H.-B.D.); the National Natural Science Foundation of China (grant nos. 12574196 to Y.Z., 12574068 to C.C. and 12504162 to H.-B.D.); the Guangdong Provincial Key Laboratory of Advanced Thermoelectric Materials and Device Physics (grant no. 2024B1212010001 to Y.Z.); the Guangdong Basic and Applied Basic Research Foundation (grant nos. 2022B1515020046, 2022B1515130005 and 2021B1515130007 to C.C.); and the Guangdong Major Project of Basic Research (grant no. 2026B0303000004 to C.C.).

**Author contributions**

Y.Z. conceived the project and designed the STM experiments. Z.Y. and S.Z. performed the STM/STS measurements and analyzed the STM data under the supervision of Y.Z., with technical support from H.-B.D. and input from J.-X.Y. Z.Y., S.Z. and Y.Z. developed the phase-resolved Fourier analysis and interpreted the CDW hierarchy. C.L. synthesized and characterized the single crystals under the supervision of G.-H.C. F.Z. performed supporting ARPES measurements under the supervision of C.C. Z.S. and X.C. performed the theoretical calculations under the supervision of J.L. C.C and G.-H.C. contributed to discussions of sample properties, electronic structure and doping-dependent band alignment. Y.Z. and Z.Y. wrote the manuscript with input from all authors. All authors discussed the results and contributed to the manuscript.

**Competing interests**

The authors declare no competing interests.

**Additional information**

Supplementary information is available for this paper.

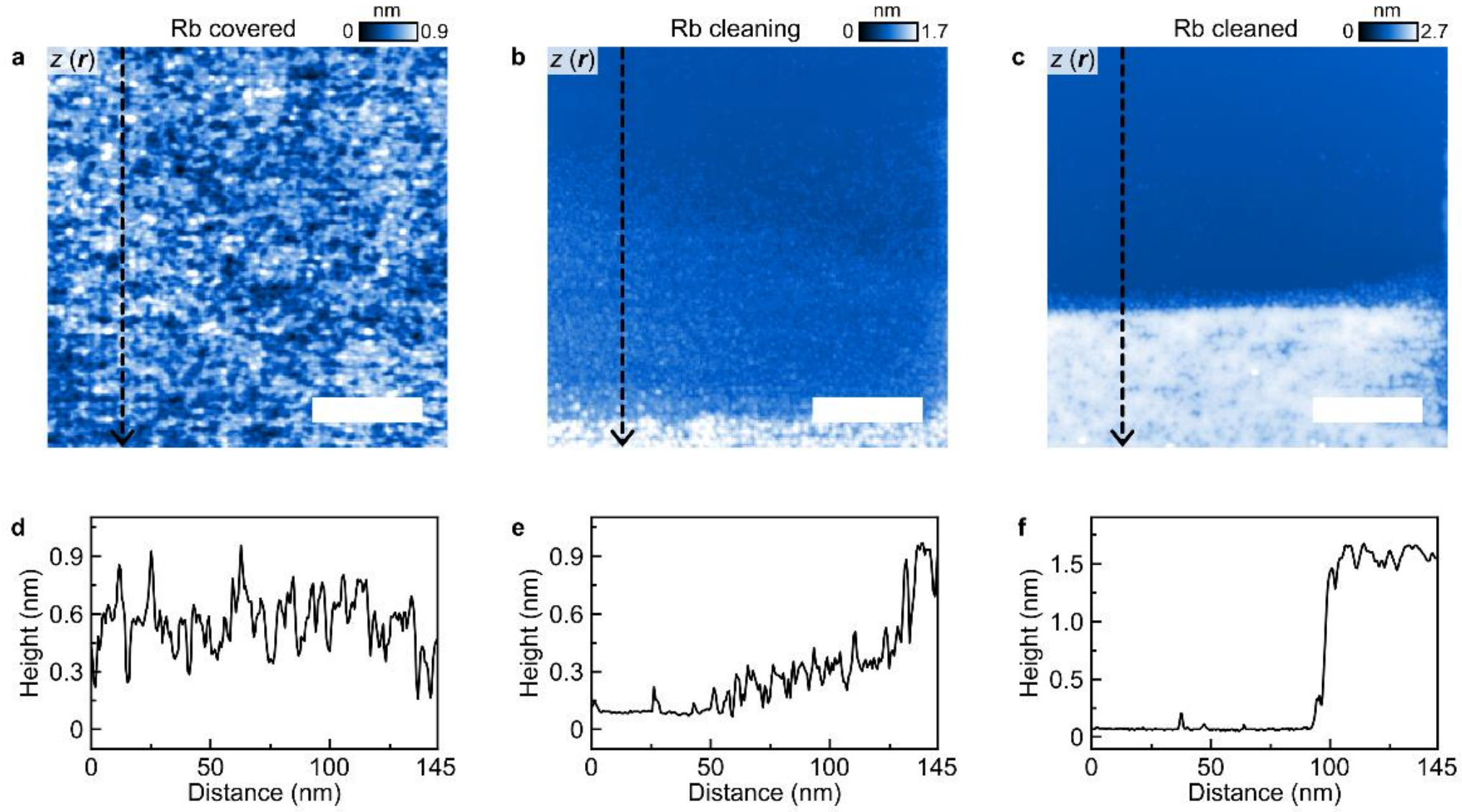

**Extended Data Fig. 1. STM preparation of the Te-terminated surface.**

**a-c,** Topographic images of the cleaved $Rb_{1-\delta}V_2Te_2O$ surface before (**a**), during (**b**) and after (**c**) in-situ surface cleaning. Scale bars: 40 nm. Setpoints: $V_{bias}$= 1 V, $I_t$ = 10 pA for **a** and **c**; $V_{bias}$ = 500 mV, $I_t$ = 10 pA for **b**. **d-f**, Height profiles extracted along the dashed arrows in **a-c**, respectively.

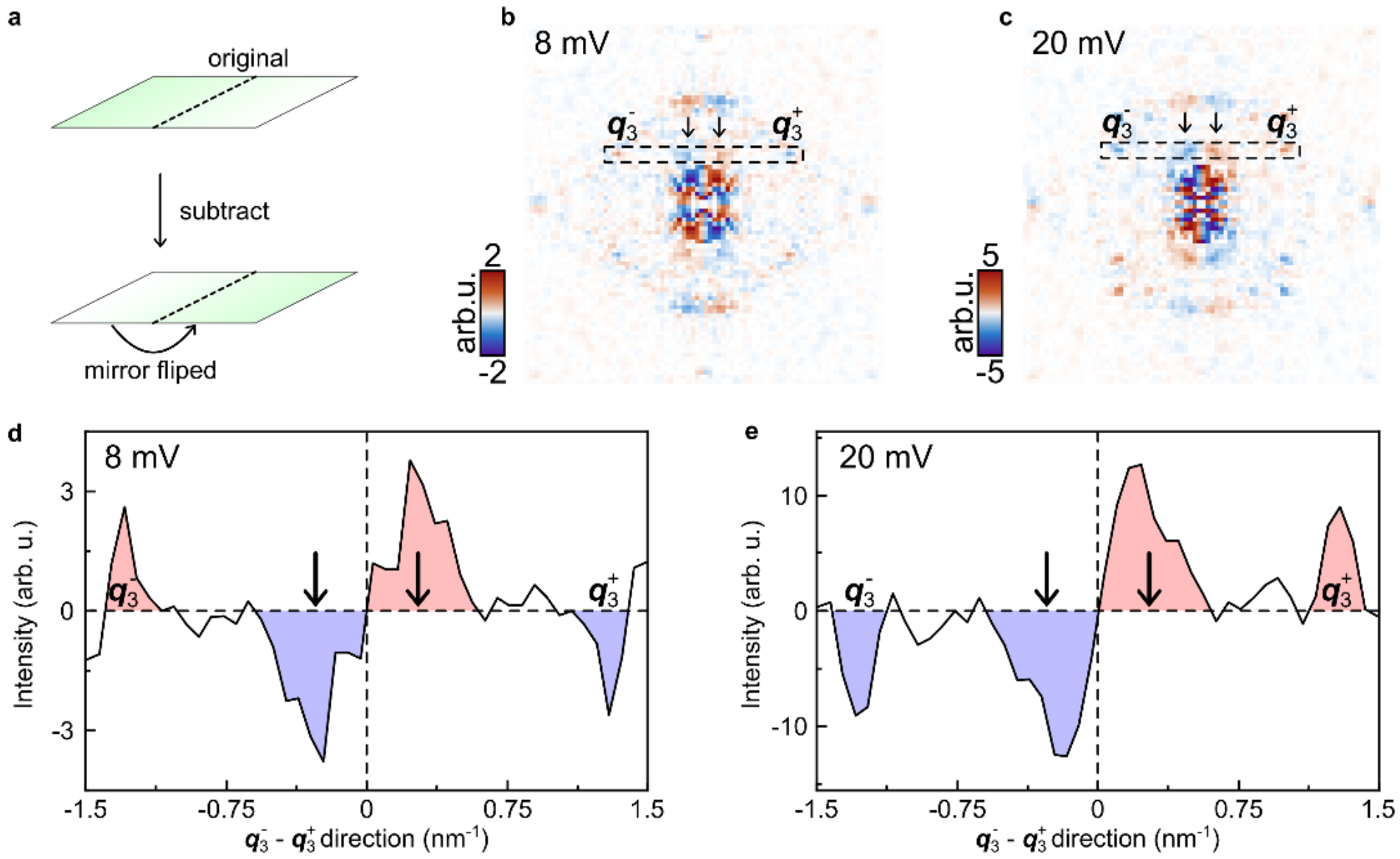

**Extended Data Fig. 2. Distinguishing the $\boldsymbol{q}_3$-specific response from mirror-odd Fourier background asymmetry.**

**a,** Schematic illustration of the mirror-antisymmetrization procedure used in **b** and **c**. For each energy, the FFT of the conductance map is subtracted from its mirror-reflected counterpart about the selected nematic axis, $\mathrm{FFT_{odd}}(\boldsymbol{k}, E) = \mathrm{FFT}(\boldsymbol{k}, E) - M\,\mathrm{FFT}(\boldsymbol{k}, E)$, to isolate mirror-odd directional asymmetry. **b, c,** Difference FFT maps $\mathrm{FFT_{odd}}(\boldsymbol{k}, E)$ acquired at 8 mV (**b**) and 20 mV (**c**). **d, e,** Line profiles extracted along $\boldsymbol{q}_3^- \rightarrow \boldsymbol{q}_3^+$ direction, marked by dashed black rectangles in **b** and **c,** at 8 mV (**d**) and 20 mV (**e**). The broad background asymmetry, indicated by black arrows, remains nearly unchanged between the two energies, whereas the $\boldsymbol{q}_3$-peak contribution reverses sign. This opposite energy evolution shows that the $\boldsymbol{q}_3$ contrast inversion does not arise from a mirror-odd background imbalance.

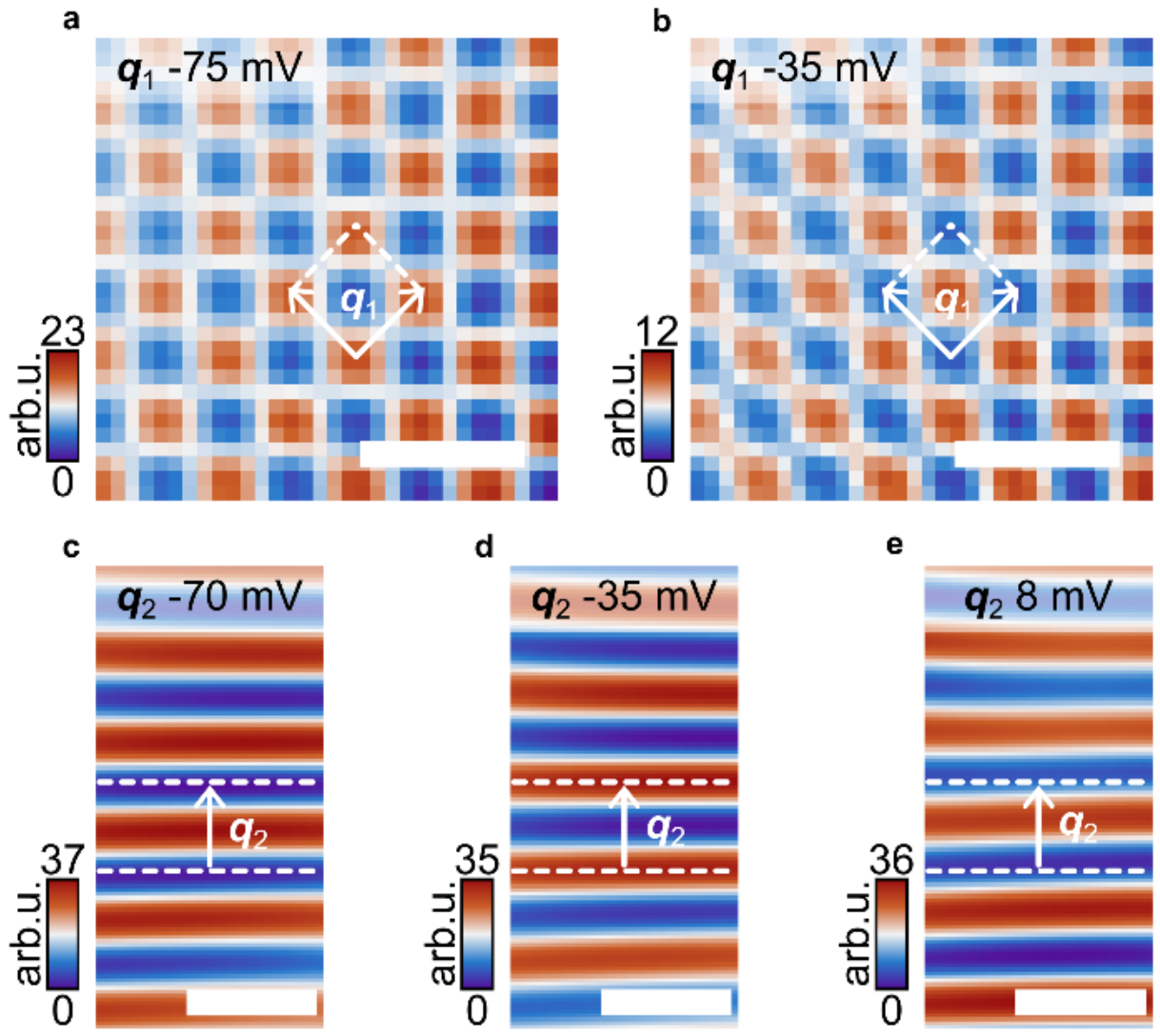

**Extended Data Fig. 3. Real-space contrast inversion of the $q_1$ and $q_2$ CDW components.**
**a, b,** Filtered real-space $\boldsymbol{q}_1$ maps at -75 mV (**a**) and -35 mV (**b**). **c-e,** corresponding $\boldsymbol{q}_2$ maps at -70 mV (**c**), -35 mV (**d**) and 8 mV (**e**). All maps were obtained by inverse FFT filtering of the corresponding $dI/dV$ maps. White arrows and dashed lines mark identical spatial positions for comparison between different bias voltages. The $\boldsymbol{q}_1$ component exhibits a clear contrast inversion between -75 and -35 mV, while the $\boldsymbol{q}_2$ component shows two contrast reversals across the displayed maps at -70, -35 and 8 mV. These real-space contrast inversions are consistent with the near π-phase shifts identified in Fig. 3. Scale bars: 1 nm in **a, b** and 3 nm in **c-e**. Setpoints: $V_{\mathrm{bias}}$ = -80 mV, $I_{\mathrm{t}}$ = 1 nA for **a-d**; $V_{\mathrm{bias}}$ = -50 mV, $I_{\mathrm{t}}$ = 500 pA for **e**.

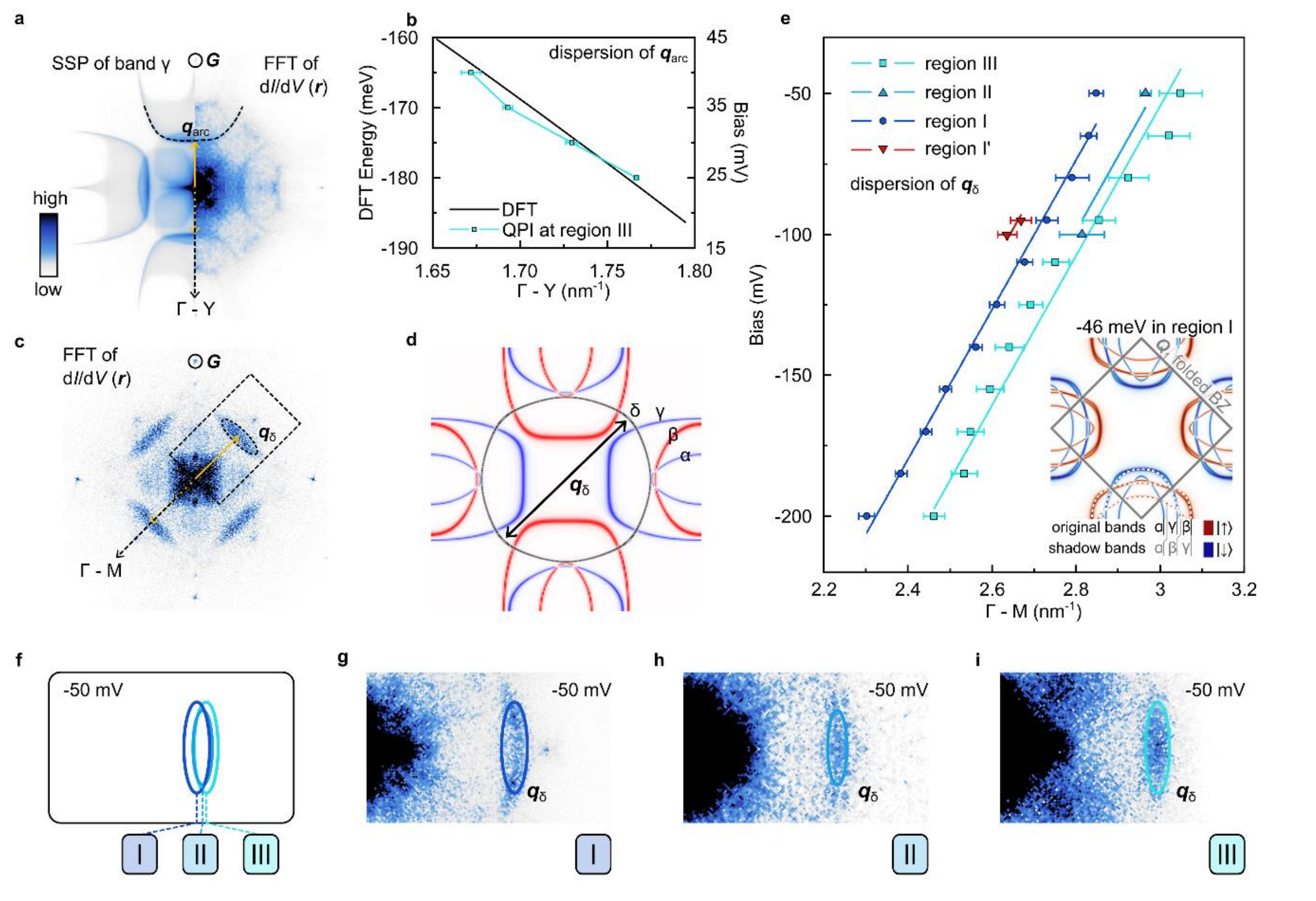

**Extended Data Fig. 4. Fermi-level determination and local doping calibration.**
**a,** Comparison between the calculated spin-selective scattering probability (SSP) pattern of the $\gamma$ band and the QPI pattern measured in region III. The arc-like feature $\boldsymbol{q}_{arc}$ matches the calculated SSP at -180 meV. Setpoint: $V_{\text{bias}}$ = 25 mV, $I_{\text{t}}$ = 100 pA. **b,** Dispersion of $\boldsymbol{q}_{arc}$ along Γ-Y direction, compared with the shifted DFT $\gamma$-band-dispersion, yielding a local Fermi-level alignment of approximately -205 meV in the DFT reference. **c,** QPI pattern measured in region I at -140 meV, showing the $\boldsymbol{q}_{\delta}$ scattering features used for relative doping calibration. Setpoint: $V_{\text{bias}}$ = -140 mV, $I_{\text{t}}$ = 200 pA. **d,** Schematic of the $\boldsymbol{q}_{\delta}$ scattering process associated with the δ band. **e,** Dispersion of the $2\boldsymbol{q}_{\delta}$ separation along the Γ-M direction across regions I-III and I'. Inset: calculated $\boldsymbol{q}_{1}$-folded constant-energy contour at -46 meV in region I, showing substantial spin-conserving overlap between the folded and original contours. **f,** Schematic of the relative δ-band scattering-pattern shifts across regions I-III. **g-i,** Cropped QPI patterns acquired in regions I-III, with $\boldsymbol{q}_{\delta}$ features highlighted by blue ellipses. QPI patterns in **a** and **g-i** are rotationally and mirror symmetrized. Setpoints: $V_{\text{bias}}$ = -50 mV (**g–i**); $I_{\text{t}}$ = 200 pA (**g**), and 100 pA (**h, i**). Error bars in **b** and **e** represent the 3σ uncertainty of peak positions extracted from Gaussian fits.

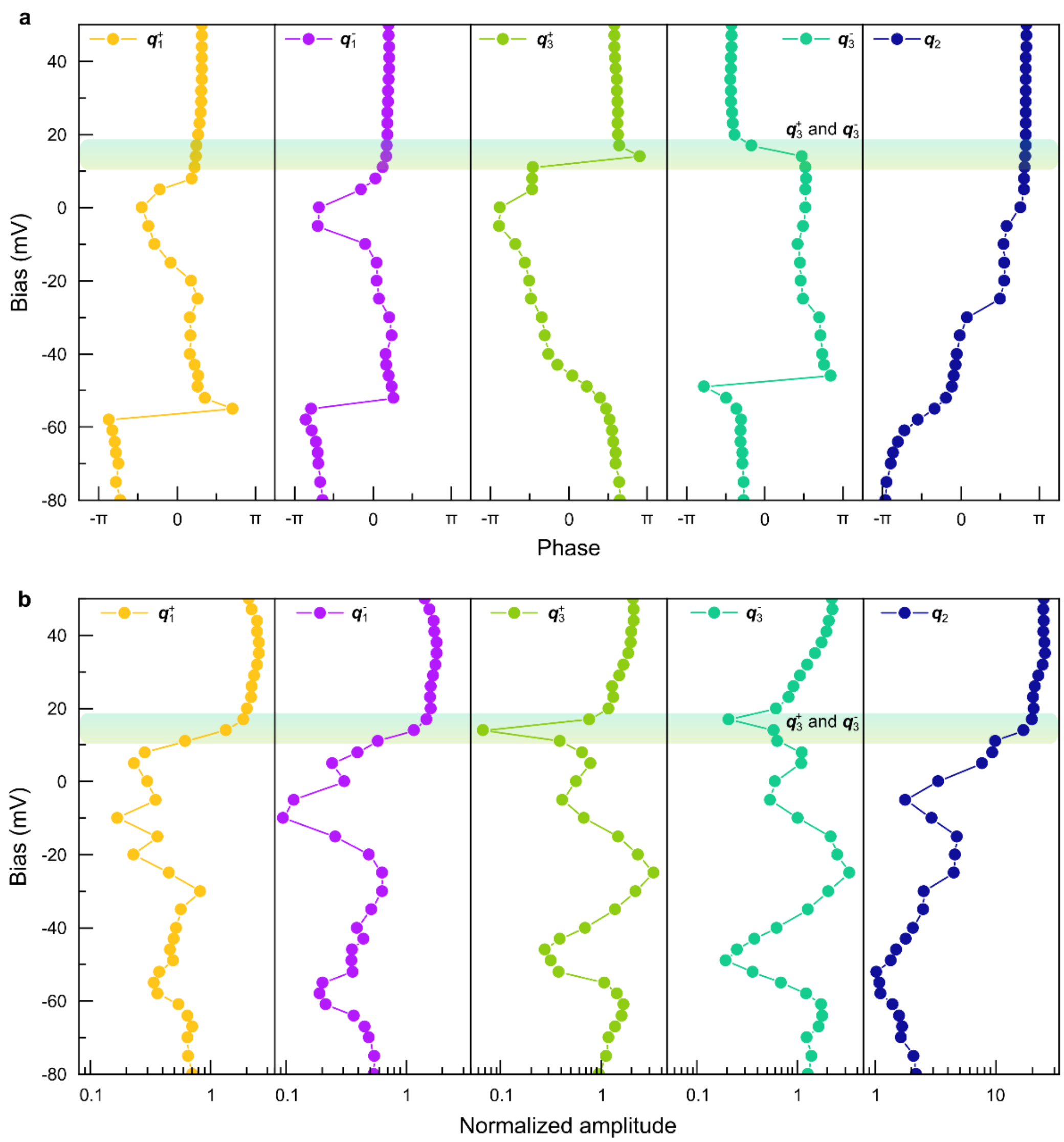

**Extended Data Fig. 5. Experimental phase and normalized-amplitude evolution of the CDW components.**

**a,** Raw Fourier phases of the CDW components, $\boldsymbol{q}_1^+$, $\boldsymbol{q}_1^-$, $\boldsymbol{q}_3^+$, $\boldsymbol{q}_3^-$ and $\boldsymbol{q}_2$, extracted from the complex FFT signal at the peak position of each ordering vector. Apparent $2\pi$ discontinuities arise from wrapping of the complex Fourier phase. **b,** Normalized FFT amplitude of each CDW component in **a**. For each ordering vector, the amplitude $A(\boldsymbol{q})$ is extracted from the FFT peak amplitude at the corresponding reciprocal-space position and normalized by the average Bragg amplitude, $[A(\boldsymbol{G}^+) + A(\boldsymbol{G}^-)]/2$. The $\boldsymbol{q}_1^+$ and $\boldsymbol{q}_1^-$ branches provide a useful comparison for assessing branch-to-branch variations. Although geometric coupling among $\boldsymbol{q}_{1,2,3}$ permits passive mixing, the pronounced redistribution between $\boldsymbol{q}_3^+$ and $\boldsymbol{q}_3^-$, together with the $\boldsymbol{q}_3$-specific amplitude suppression and phase reversal, occurs in an energy range where the $\boldsymbol{q}_1^+$, $\boldsymbol{q}_1^-$, and $\boldsymbol{q}_2$ phases vary weakly and normalized amplitudes stay increasing, indicating that passive mixing alone is insufficient to account for the branch-resolved evolution of the off-axis $\boldsymbol{q}_3$ response.

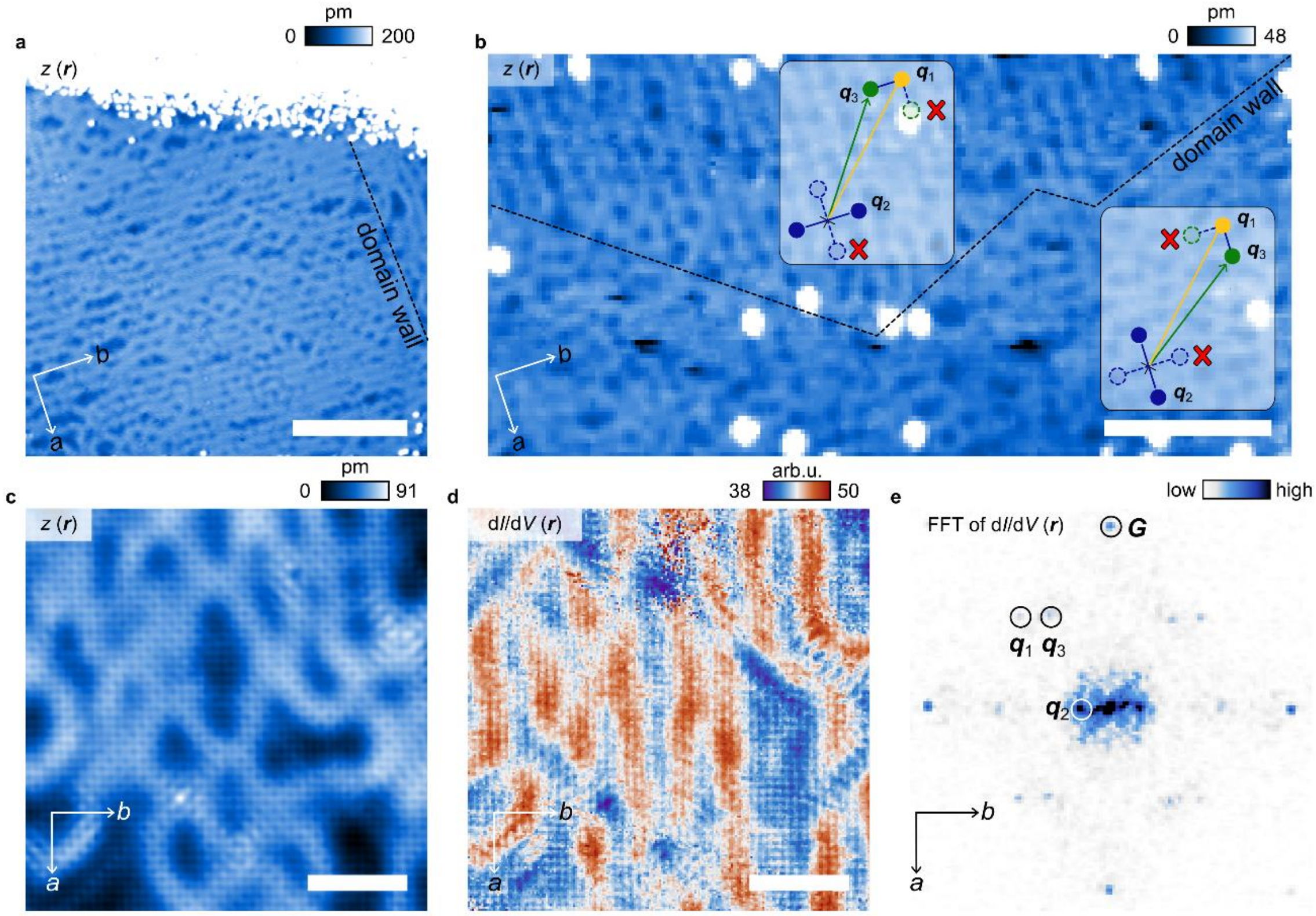

**Extended Data Fig. 6. Domain-selective development of the $q_2$ and $q_3$ orders.**
**a, b,** Topographic images showing spatial domains of the $\boldsymbol{q}_2$ order. Domain boundaries are marked by dashed black lines. Insets display schematics of the corresponding CDW vectors in each domain. Scale bars: 20 nm in **a** and 15 nm in **b**. Setpoints: $V_{\text{bias}}$ = 100 mV, $I_{\text{t}}$ = 10 pA for **a**; $V_{\text{bias}}$ = 800 mV, $I_{\text{t}}$ = 10 pA for **b**. **c,** Topographic image of a region exhibiting anisotropic and partially developed $\boldsymbol{q}_2$ order along the ***a*** axis, distinct from the ***b***-axis-oriented $\boldsymbol{q}_2$ order shown in Fig. 1. Setpoint: $V_{\text{bias}}$ = 40 mV, $I_{\text{t}}$ = 500 pA. **d,** Corresponding d*I*/d*V* map of the same region shown in **c**, measured at 44 mV. Stripe-like modulation of the local density of states is observed along the ***a*** direction. Setpoint: $V_{\text{bias}}$ = 44 mV, $I_{\text{t}}$ = 400 pA. **e,** FFT of the d*I*/d*V* map in **d**. The $\boldsymbol{q}_3$ peak, marked by the black circles, appears in the symmetry-related sector opposite to that shown in Fig. 1. The reversal of the $\boldsymbol{q}_3$ orientation together with the switching of the $\boldsymbol{q}_2$ domain indicates that the $\boldsymbol{q}_3$ sector is locked to the local nematic environment selected by $\boldsymbol{q}_2$. Scale bars: 5 nm in **c**, **d**.

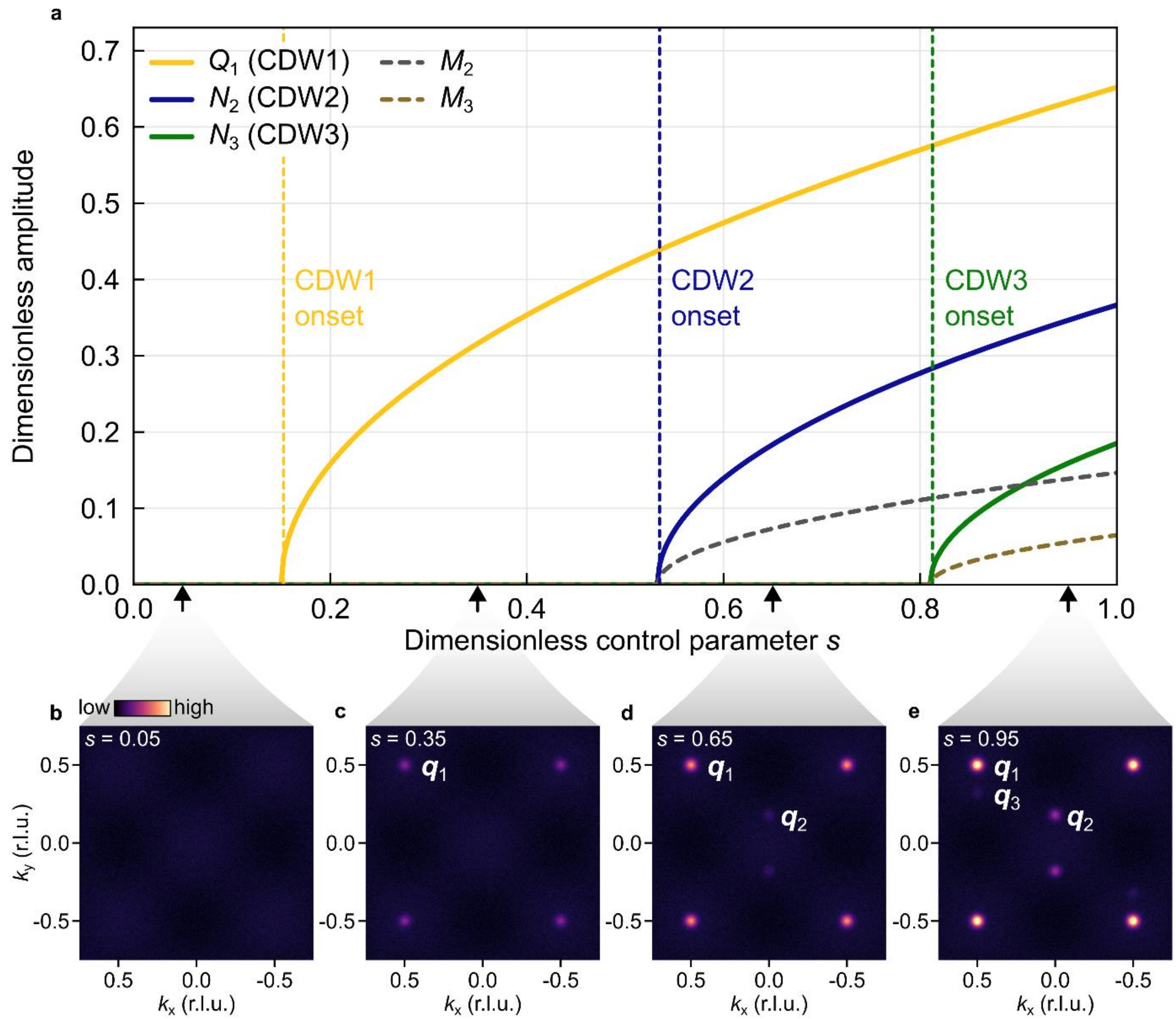

**Extended Data Fig. 7. Phenomenological Landau cascade and synthetic FFT peak maps.**
**a,** Schematic Landau cascade showing the dimensionless amplitudes of the CDW order parameters and octupole components as a function of the dimensionless control parameter $s$. Solid lines denote the amplitude of $Q_1$, $N_2$, and $N_3$. Dashed lines denote the octupole components $M_2$ and $M_3$, which are induced concurrently with $N_2$, and $N_3$, respectively. Black arrows indicate the values of $s$ for generating the synthetic FFT peak maps in **b-e**. **b-e,** Synthetic FFT peak maps generated from the Landau amplitudes at the selected values of $s$, illustrating the progressive emergence of $\boldsymbol{q}_1$, $\boldsymbol{q}_2$, and $\boldsymbol{q}_3$ components. The phenomenological construction is described in the Numerical realization section of the Supplementary Information.